\documentclass{aastex61}

\usepackage{longtable}
\usepackage{multirow}

\received{\today}
\submitjournal{ApJ}

\shorttitle{An ALMA view of the N159E-Papillon Nebula}
\shortauthors{Fukui et al.}

\begin{document}

\title{An ALMA view of molecular filaments in the Large Magellanic Cloud. I. The formation of high-mass stars and pillars in the N159E-Papillon Nebula triggered by a cloud--cloud collision}

\correspondingauthor{Yasuo Fukui}
\email{fukui@a.phys.nagoya-u.ac.jp}

\author{Yasuo Fukui}
\affiliation{Department of Physics, Nagoya University, Chikusa-ku, Nagoya 464-8602, Japan}
\affiliation{Institute for Advanced Research, Nagoya University, Furo-cho, Chikusa-ku, Nagoya 464-8601, Japan}

\author[0000-0002-2062-1600]{Kazuki Tokuda}
\affiliation{Department of Physical Science, Graduate School of Science, Osaka Prefecture University, 1-1 Gakuen-cho, Naka-ku, Sakai, Osaka 599-8531, Japan}
\affiliation{National Astronomical Observatory of Japan, National Institutes of Natural Science, 2-21-1 Osawa, Mitaka, Tokyo 181-8588, Japan}

\author{Kazuya Saigo}
\affiliation{National Astronomical Observatory of Japan, National Institutes of Natural Science, 2-21-1 Osawa, Mitaka, Tokyo 181-8588, Japan}

\author{Ryohei Harada}
\affiliation{Department of Physical Science, Graduate School of Science, Osaka Prefecture University, 1-1 Gakuen-cho, Naka-ku, Sakai, Osaka 599-8531, Japan}

\author{Kengo Tachihara}
\affiliation{Department of Physics, Nagoya University, Chikusa-ku, Nagoya 464-8602, Japan}

\author{Kisetsu Tsuge}
\affiliation{Department of Physics, Nagoya University, Chikusa-ku, Nagoya 464-8602, Japan}

\author{Tsuyoshi Inoue}
\affiliation{Department of Physics, Nagoya University, Chikusa-ku, Nagoya 464-8602, Japan}

\author{Kazufumi Torii}
\affiliation{Nobeyama Radio Observatory, 462-2 Nobeyama Minamimaki-mura, Minamisaku-gun, Nagano 384-1305, Japan}

\author{Atsushi Nishimura}
\affiliation{Department of Physical Science, Graduate School of Science, Osaka Prefecture University, 1-1 Gakuen-cho, Naka-ku, Sakai, Osaka 599-8531, Japan}

\author{Sarolta Zahorecz}
\affiliation{Department of Physical Science, Graduate School of Science, Osaka Prefecture University, 1-1 Gakuen-cho, Naka-ku, Sakai, Osaka 599-8531, Japan}
\affiliation{National Astronomical Observatory of Japan, National Institutes of Natural Science, 2-21-1 Osawa, Mitaka, Tokyo 181-8588, Japan}

\author{Omnarayani Nayak}
\affiliation{Space Telescope Science Institute, 3700 San Martin Drive, Baltimore, MD 21218, USA}

\author{Margaret Meixner}
\affiliation{Department of Physics \& Astronomy, Johns Hopkins University, 3400 N. Charles Street, Baltimore, MD 21218, USA}
\affiliation{Space Telescope Science Institute, 3700 San Martin Drive, Baltimore, MD 21218, USA}

\author{Tetsuhiro Minamidani}
\affiliation{Nobeyama Radio Observatory, 462-2 Nobeyama Minamimaki-mura, Minamisaku-gun, Nagano 384-1305, Japan}

\author{Akiko Kawamura}
\affiliation{National Astronomical Observatory of Japan, National Institutes of Natural Science, 2-21-1 Osawa, Mitaka, Tokyo 181-8588, Japan}

\author{Norikazu Mizuno}
\affiliation{National Astronomical Observatory of Japan, National Institutes of Natural Science, 2-21-1 Osawa, Mitaka, Tokyo 181-8588, Japan}
\affiliation{Department of Astronomy, School of Science, The University of Tokyo, 7-3-1 Hongo, Bunkyo-ku, Tokyo 133-0033, Japan}

\author{Remy Indebetouw}
\affiliation{Department of Astronomy, University of Virginia, P.O. Box 400325, Charlottesville, VA 22904, USA}
\affiliation{National Radio Astronomy Observatory, 520 Edgemont Road, Charlottesville, VA 22903, USA}

\author{Marta Sewi{\l}o}
\affiliation{CRESST II and Exoplanets and Stellar Astrophysics Laboratory, NASA, Goddard Space Flight Center, Greenbelt, MD 20771, USA}
\affiliation{Department of Astronomy, University of Maryland, College Park, MD 20742, USA}

\author{Suzanne Madden}
\affiliation{AIM, CEA, CNRS, Universit\'e Paris-Saclay, Universit\'e Paris Diderot, Sorbonne Paris Cit\'e, F-91191 Gif-sur-Yvette, France}

\author{Maud Galametz}
\affiliation{AIM, CEA, CNRS, Universit\'e Paris-Saclay, Universit\'e Paris Diderot, Sorbonne Paris Cit\'e, F-91191 Gif-sur-Yvette, France}

\author{Vianney Lebouteiller}
\affiliation{AIM, CEA, CNRS, Universit\'e Paris-Saclay, Universit\'e Paris Diderot, Sorbonne Paris Cit\'e, F-91191 Gif-sur-Yvette, France}

\author{C.-H. Rosie Chen}
\affiliation{Max Planck Institute for Radio Astronomy, Auf dem Huegel 69, D-53121 Bonn, Germany}

\author{Toshikazu Onishi}
\affiliation{Department of Physical Science, Graduate School of Science, Osaka Prefecture University, 1-1 Gakuen-cho, Naka-ku, Sakai, Osaka 599-8531, Japan}



\begin{abstract}
We present the ALMA observations of CO isotopes and 1.3\,mm continuum emission toward the N159E-Papillon Nebula in the Large Magellanic Cloud (LMC). The spatial resolution is 0\farcs25--0\farcs28 (0.06--0.07\,pc), which is a factor of 3 higher than the previous ALMA observations in this region. The high resolution allowed us to resolve highly filamentary CO distributions with typical widths of $\sim$0.1\,pc (full width half maximum) and line masses of a few 100\,$M_{\odot}$\,pc$^{-1}$. The filaments (more than ten in number) show an outstanding hub-filament structure emanating from the nebular center toward the north. We identified for the first time two massive protostellar outflows of $\sim$10$^4$ yr dynamical age along one of the most massive filaments. The observations also revealed several pillar-like CO features around the Nebula. The H$\;${\sc ii} region and the pillars have a complementary spatial distribution and the column density of the pillars is an order of magnitude higher than that of the pillars in the Eagle nebula (M16) in the Galaxy, suggesting an early stage of pillar formation with an age younger than $\sim$10$^5$ yr. We suggest that a cloud--cloud collision triggered the formation of the filaments and protostar within the last $\sim$2\,Myr. 
It is possible that the collision is more recent, as part of the kpc-scale H$\;${\sc i} flows come from the tidal interaction resulting from the close encounter between the LMC and SMC $\sim$200\,Myr ago as suggested for R136 by Fukui et al.

\end{abstract}


\section{Introduction} \label{sec:intro}
Obtaining a comprehensive picture of high-mass star formation is one of the most important issues in astrophysics. Although several promising mechanisms of high-mass star formation have been proposed and explored \citep[e.g.,][]{Zinnecker07,Tan14}, these theories are not yet able to fully explain various aspects of high-mass star formation through when compared with observations. We note that many isolated or asymmetric O-star formation (e.g., M42, \citealt{Fukui18a}; GM24, \citealt{Fukui18b}; RCW34, \citealt{Hayashi18}; see also \citealt{Ascenso18}) and the production of dense and massive cores as the initial condition of high-mass star formation as posited in theories \citep[e.g.,][]{Krumholz09}, for instance, remain important challenges to be addressed by the various star formation theories.

Most recently, supersonic cloud--cloud collisions (i.e., gas compression by collisions between different-velocity gas streams) have been extensively explored and tested as a trigger of high-mass star formation both by theoretical and observational studies (see the PASJ special issue: Star Formation Triggered by Cloud--Cloud Collision (Volume 70, Issue SP2) and the other papers). Among them, the magnetohydrodynamic (MHD) numerical simulations made by \cite{Inoue18} showed that massive filaments with high-mass stars at their vertex are naturally formed at the shock-compressed layer by a cloud--cloud collision (see also \citealt{Inoue13}). According to their simulations, there is no significant delay between the formation of the filament and the first sink particle (the high-mass protostar)---more than a few 0.1\,Myr. Although the intrinsic turbulent motions in a molecular cloud form filamentary structures without any colliding flows, more massive/high-density hub-filamentary structures \citep[see][]{Myers09} tend to be created preferentially in the colliding case, as supported by \cite{Matsumoto15}, \cite{Balfour15}, \cite{Wu17}, and \cite{Takahira18}. Hub filaments associated with high-mass protostellar sources are often seen in many Galactic high-mass star-forming regions, such as infrared dark clouds \citep[e.g.,][]{Busquet13,Peretto13,Williams18}.

Concerning observational aspects, multiple components of molecular clouds are often seen as a possible evidence of cloud--cloud collisions in the Galactic cluster-forming regions \citep[e.g.,][]{Furukawa09,Fukui16,Nishimura18}. In addition, recent observations toward high-mass star-forming regions with a  resolution higher than 0.1\,pc have been attempting to resolve filamentary structures, which may be crucial in regulating star formation efficiency (e.g., NGC\,6334, \citealt{Andre16}). Even though the filament width of high-mass star-forming regions is not significantly different from the typical filament width of $\sim$0.1\,pc in the low-mass star-forming regions \citep{Andre14}, the line mass (mass per unit length) of filaments is one or two orders of magnitude higher than the critical isothermal line mass \citep[see][]{Inutsuka97}. \cite{Andre16} suggests that the massive filaments that we see in the present day may be an indirect piece of evidence for their very recent formation because such super critical filaments are supposed to be unstable against the gravitational collapse. This suggestion is consistent with the theory that massive filaments and high-mass stars are rapidly formed by shock compression that originated in cloud--cloud collisions or converging flows.

Observations toward external galaxies can be very powerful in the study of high-mass star formation studies compared to those in the Milky Way because we can avoid the serious contaminations along the lines of sight toward the galactic plane. However, larger distances generally make it difficult to achieve a spatial resolution high enough to resolve filaments. In this regard, the Large Magellanic Cloud (LMC) is close to us at a distance of $\sim$50\,kpc \citep{Schaefer08,de14}, and its face-on view \citep{Balbinot15} allows us to isolate associations among the molecular clouds, massive YSOs, and  H$\;${\sc ii} regions without the abovementioned contamination problem. Extensive molecular gas surveys toward the LMC identified $\sim$300 giant molecular clouds (GMCs) \citep{Fukui99,Fukui08,Kawamura09,Fukui10} and found that the N159 GMC is a crucial object as the most active ongoing star formation site among the samples.

Our present target, the N159E-Papillon region, is an ideal source to investigate the filament/high-mass star formation as well as the feedback from massive YSOs. The H$\alpha$ observations with \textit{Hubble Space Telescope} discovered a butterfly shaped  H$\;${\sc ii} region, $``$the Papillon Nebula,$"$ with the wings separated by $\sim$0.6\,pc \citep{Heydari99,Meynadier04}. The very compact size of the  H$\;${\sc ii} region indicates that the embedded YSO system (hereafter, the Papillon Nebula YSO), which contains two massive protostars with the stellar masses of $\sim$21\,$M_{\odot}$ \citep{Heydari99,Testor07} and $\sim$41\,$M_{\odot}$ \citep{Indebetouw04,Chen10}, is in a very young stage of the onset of the surrounding gas ionization. The previous ALMA Cycle 1 observations with an angular resolution of $\sim$0.24\,pc (\citealt{Saigo17}, hereafter Paper I) found that at least three filamentary CO clouds with different velocities are entangled toward the Papillon Nebula, and that the shape of the Papillon Nebula nicely fits the cavity of CO emission. Based on these results, the authors suggested that a filamentary cloud collision triggered high-mass star formation, analogous to N159W.
Prior to N159E, \cite{Fukui15} made a similar analysis of the ALMA Cycle 1 data in N159W and found that two CO filaments are crossed with each other and that a high-mass protostar showing outflow is located toward the crossing position. This was the first protostellar outflow discovered outside the Milky Way, lending further support that protostellar outflows are ubiquitous in protostellar evolution in galaxies. 
The N159E and N159W regions are separated by $\sim$50\,pc in the sky, which requires 10\,Myr to travel at a typical velocity dispersion of $\sim$5 km\,s$^{-1}$, longer than the protostellar age of 1--5\,Myr. However, it remained somewhat curious that the extension of the filamentary clouds is seen toward the same direction in fan shape, to the north of the young stars, both in N159E and N159W (see also Figure 6 of Paper I).

In this paper, we present new ALMA Cycle 4 observations toward the N159E-Papillon region with a factor of three higher angular resolution than the previous study. Section \ref{sec:Obs} contains descriptions of the observations and in Section \ref{Results} containes the results, where 
we mainly focus on the filamentary molecular clouds and the pillar-like structure around the Papillon Nebula. In addition, YSOs are newly identified, with indications of outflow wings along the filamentary structure. In Sect. \ref{Dis}, we discuss the formation scenario of molecular filaments, high-mass stars, and the pillars. Note that we also observed two other targets (the N159W-South and the N159W-North region) in this project, and the observational results of the N159W-South region are presented in a separate paper (\citealt{Tokuda19}, hereafter, TFH19).

\section{Observations} \label{sec:Obs}
We conducted ALMA observations toward the N159E-Papillon region with Band 6 (211--275 GHz) in Cycle 4 as a part of the multi-objects survey in N159 (P.I. Y. Fukui \#2016.1.01173.S.) by using the 12 m array alone. The central coordinate of the observed field toward the N159E-Papillon region was ($\alpha_{\rm J2000.0}$, $\delta_{\rm J2000.0}$) = (5$^{\rm h}$40$^{\rm m}$04\fs0, -69\arcdeg44\arcmin34\farcs0). The target lines were $^{12}$CO\,($J$ = 2--1), $^{13}$CO\,($J$ = 2--1), C$^{18}$O\,($J$ = 2--1), SiO\,($J$ = 5--4), and H30$\alpha$. The observation settings and the data reduction processes are described in TFH19. In the imaging process, we used $``$tclean$"$, which is implemented in the Common Astronomy Software Application package \citep{McMullin07}, with a noise threshold of the $\sim$1$\sigma$ noise level. We did not apply the self-calibration process. The synthesized beam sizes of the molecular lines ($^{12}$CO, $^{13}$CO, and C$^{18}$O) are 0\farcs28\,$\times$\,0\farcs25, using the Briggs weighting with a robust parameter of 0.5. The sensitivities (RMS noise levels) of the lines are $\sim$4.0\,mJy\,beam$^{-1}$ ($\sim$1.4 K) for $^{12}$CO, $\sim$4.5\,mJy\,beam$^{-1}$ ($\sim$1.5 K) for $^{13}$CO, and $\sim$4.5\,mJy\,beam$^{-1}$ ($\sim$1.2 K) for C$^{18}$O at a velocity resolution of 0.2 km\,s$^{-1}$. The beam size and RMS noise level for the 1.3\,mm continuum map are 0\farcs26\,$\times$\,0\farcs23, and $\sim$0.027\,mJy\,beam$^{-1}$, respectively.
There is no significant missing flux in the present data set compared to our Cycle 1 observations (Paper I). The total fluxes of the $^{12}$CO, $^{13}$CO, and the 1.3\,mm continuum data of the Cycle 4 were $\sim$20\% lower than those of the Cycle 1 {(see also the detailed flux comparison between the two data sets shown in Appendix \ref{A:MissingF})}. We thus use the Cycle 4 data alone in this paper. 
Actually, the $^{13}$CO intensity distributions of the new data have reasonably reproduced the overall CO distribution obtained in our previous lower-resolution study (see also Figure \ref{fig:13COfilament} in Sect. \ref{R:13COfilament}).

We made moment-masked cube data \citep[e.g.,][]{Dame11,Nishimura15} to suppress the noise effect in the velocity analyses. We set the emission-free pixels, which are determined by significant emission from the smoothed data in the velocity/spatial axes, as zero values. We used the masked cube data to make moment maps.

\section{Results \label{Results}} 
\subsection{$^{13}$CO Filaments in the N159E-Papillon region\label{R:13COfilament}}

\begin{figure}[htbp]
\includegraphics[width=180mm]{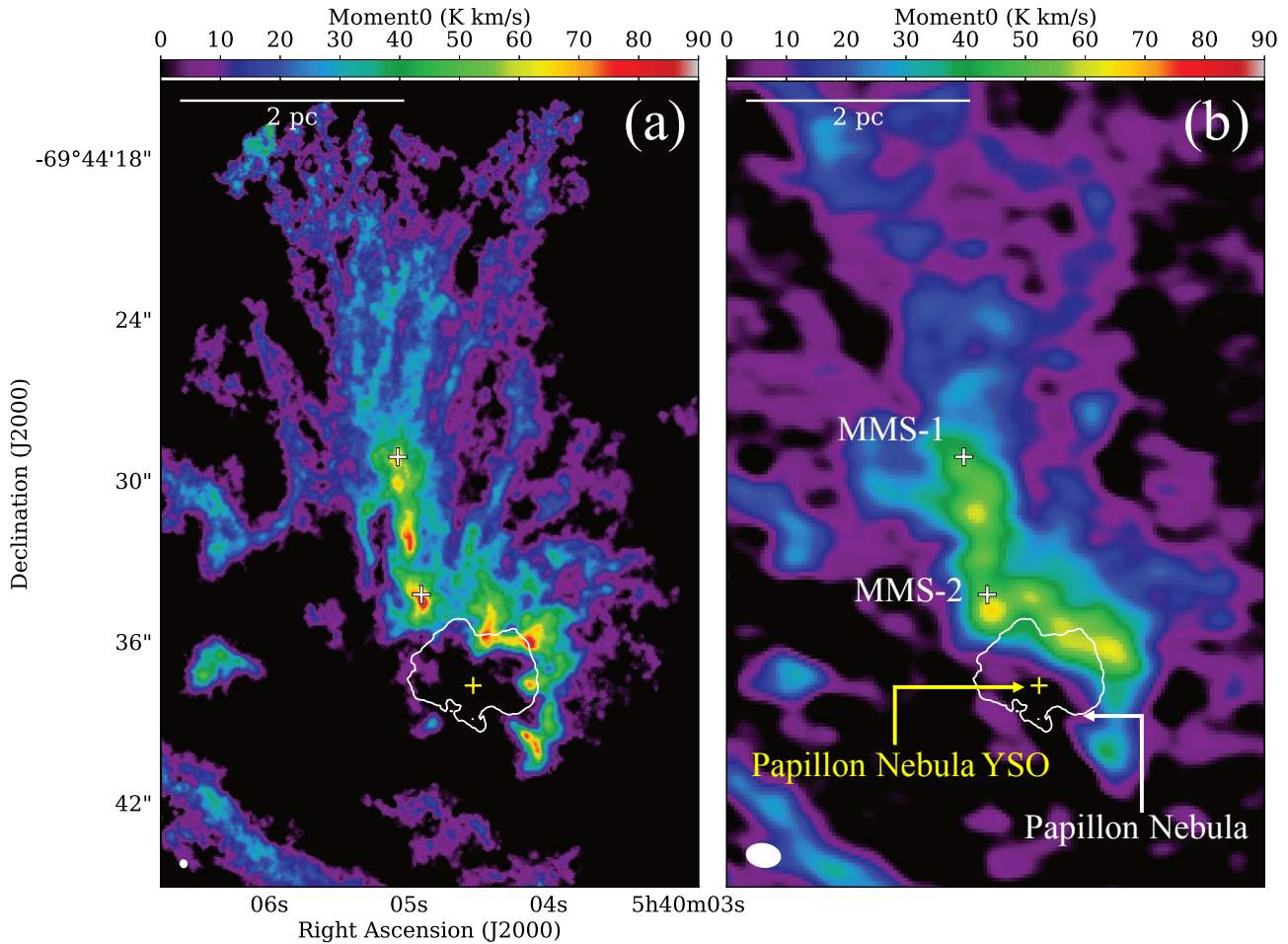}
\caption{$^{13}$CO\,($J$ = 2--1) images toward the N159E-Papillon region. (a) Velocity-integrated intensity of $^{13}$CO\,($J$ = 2--1) obtained in Cycle 4. The angular resolutions are shown in the lower left corners in each panel. The yellow crosses represent the position of the Papillon Nebula YSO, given by the 98 GHz continuum peak (Paper I). The white crosses denote the position of MMS-1 and MMS-2 (see the text in Sect. \ref{R:outflow}). The shape of the Papillon Nebula traced by H$\alpha$ emission (see also Figure \ref{fig:COpillars} (b)) is shown in white contours in both panels. (b) Same as (a) but for the $^{13}$CO\,($J$ = 2--1) data obtained in Cycle 1 (Paper I). {Note that the moment masked cube data are used in both panels (see Sect. \ref{sec:Obs}).} \label{fig:13COfilament}} 
\end{figure}

\begin{figure}[htbp]
\includegraphics[width=180mm]{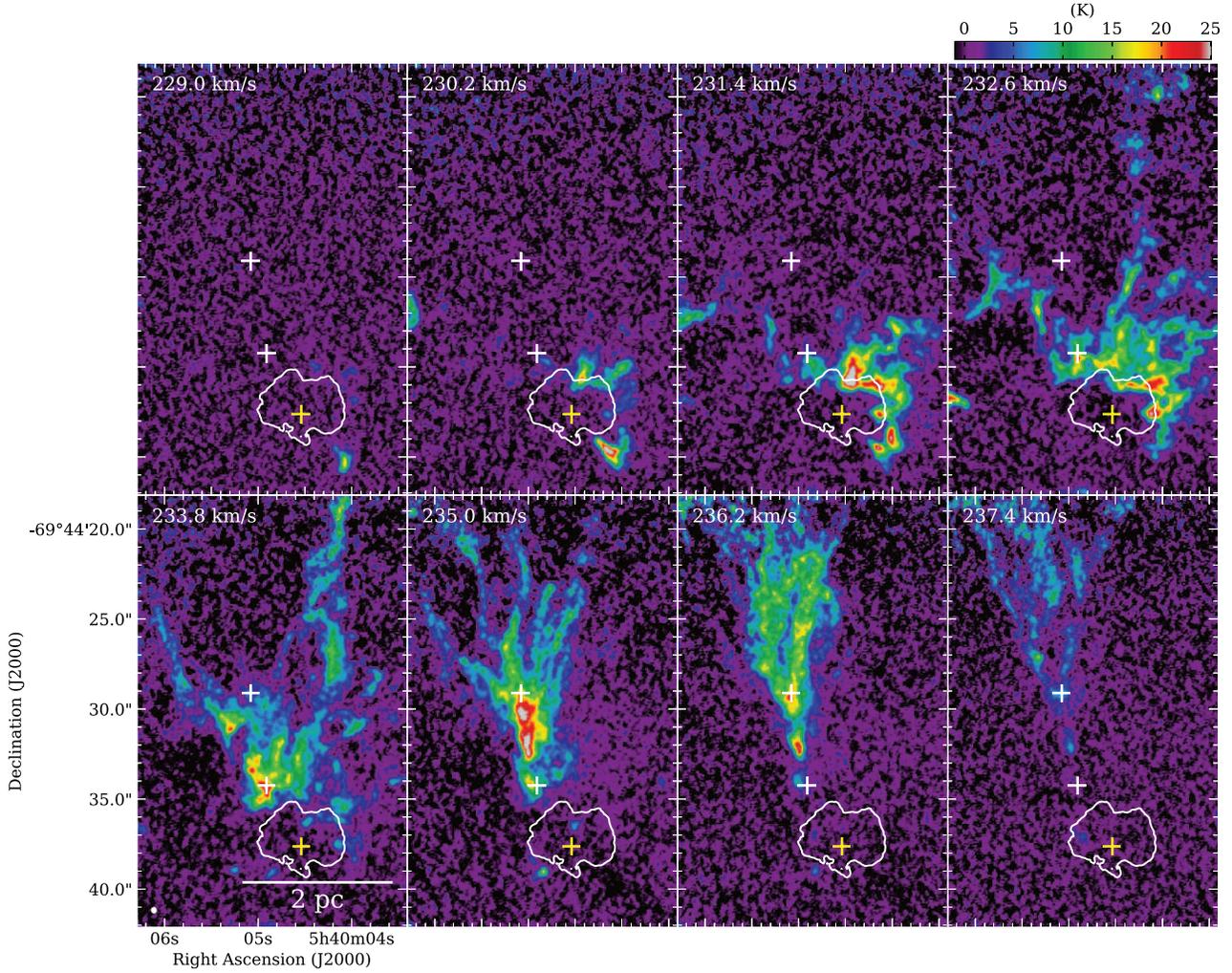}
\caption{Velocity-channel maps toward the N159E-Papillon region in $^{13}$CO\,($J$ = 2--1). The lowest velocities are given in the upper left corners in each panel. The white/yellow crosses and white lines are the same as those in Figure \ref{fig:13COfilament}. The angular resolution is given by a white ellipse in the lower left corner in the lower left panel. \label{fig:13COchanmap}}
\end{figure}

\begin{figure}[htbp]
\includegraphics[width=180mm]{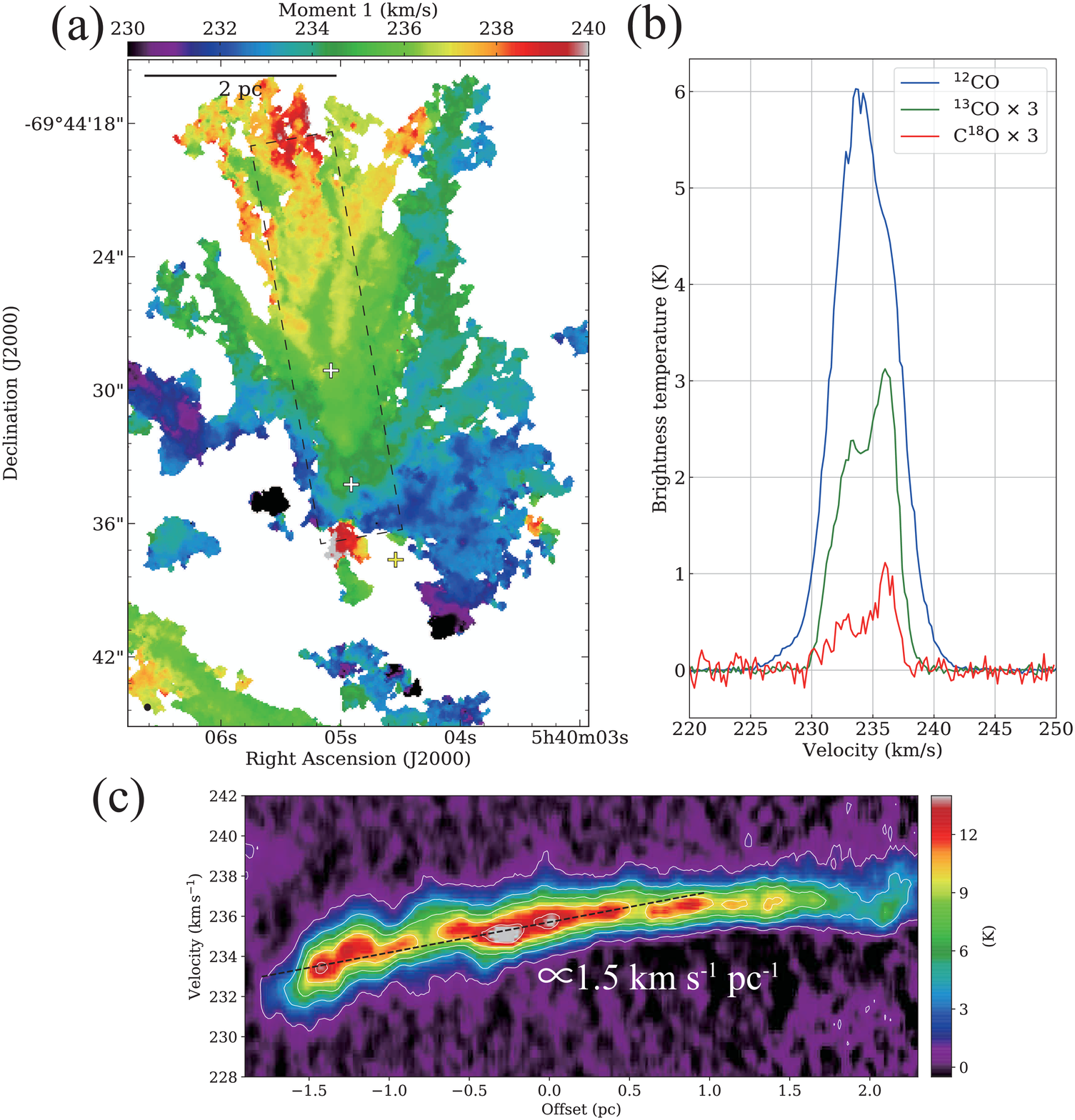}
\caption{Velocity structures of filamentary molecular clouds traced by the CO observations in the N159E-Papillon region. (a) The first-moment intensity-weighted velocity map of $^{13}$CO\,($J$ = 2--1) is shown in color scale. Note that the moment-masked cube data is used (see Section \ref{sec:Obs}). The white/yellow crosses are the same as those in Figure \ref{fig:13COfilament}. The angular resolution is shown in the lower left corner. 
(b) Averaged spectra of $^{12}$CO, $^{13}$CO, and C$^{18}$O\,($J$= 2--1) in the regions where the molecular line emissions were detected.
(c) A $^{13}$CO\,($J$ = 2--1) position-velocity diagram along the region shown by dashed rectangle in panel (a). The X-axis shows angular distances in parsec scale from the position of MMS-1. The dotted line shows the velocity gradient (1.5\,km\,s$^{-1}$\,pc$^{-1}$) derived by the least-squares fitting (see the text).} \label{fig:Moment1}
\end{figure}

Figure \ref{fig:13COfilament} shows integrated intensities of $^{13}$CO\,($J$ = 2--1) toward the N159E-Papillion region obtained in the ALMA Cycle 4 and 1 (Paper I).
We found  overall agreement with each other, while the Cycle 4 data clearly resolve significantly more details of the filaments. Figure \ref{fig:13COchanmap} shows the velocity-channel map of the region,  which also shows the multiple filamentary distributions. Most of the filaments are aligned roughly in the north--south direction. The first-moment map of $^{13}$CO\,($J$ = 2--1) in Figure \ref{fig:Moment1} (a) shows the velocity gradient along that direction. Note that a similar figure was also presented in the previous lower-resolution study (Figure 6(a) in Paper I). The averaged $^{12}$CO, $^{13}$CO, and C$^{18}$O spectra shown in Figure \ref{fig:Moment1} (b) demonstrate that there are no velocity components significantly over the blueshifted ($\lesssim$225\,km\,s$^{-1}$) and redshifted ($\gtrsim$245\,km\,s$^{-1}$) ranges. This means that there are no external components affecting the determination of the first-moment map (Figure \ref{fig:Moment1} (a)). Figure \ref{fig:Moment1} (c) shows the $^{13}$CO position-velocity (PV) diagram along the dashed rectangle in Figure \ref{fig:Moment1} (a). 
We estimated the centroid velocity in each X-axis bin with Gaussian fittings, and then the velocity gradient was derived by the least-squares fitting with a linear function within the X-range of -1.8 to 1.0\,pc. The velocity gradient is $\sim$1.5\,km\,s$^{-1}$\,pc$^{-1}$ and its possible origin will be discussed in Sect. \ref{D:filament}.

\begin{figure}[htbp]
\includegraphics[width=180mm]{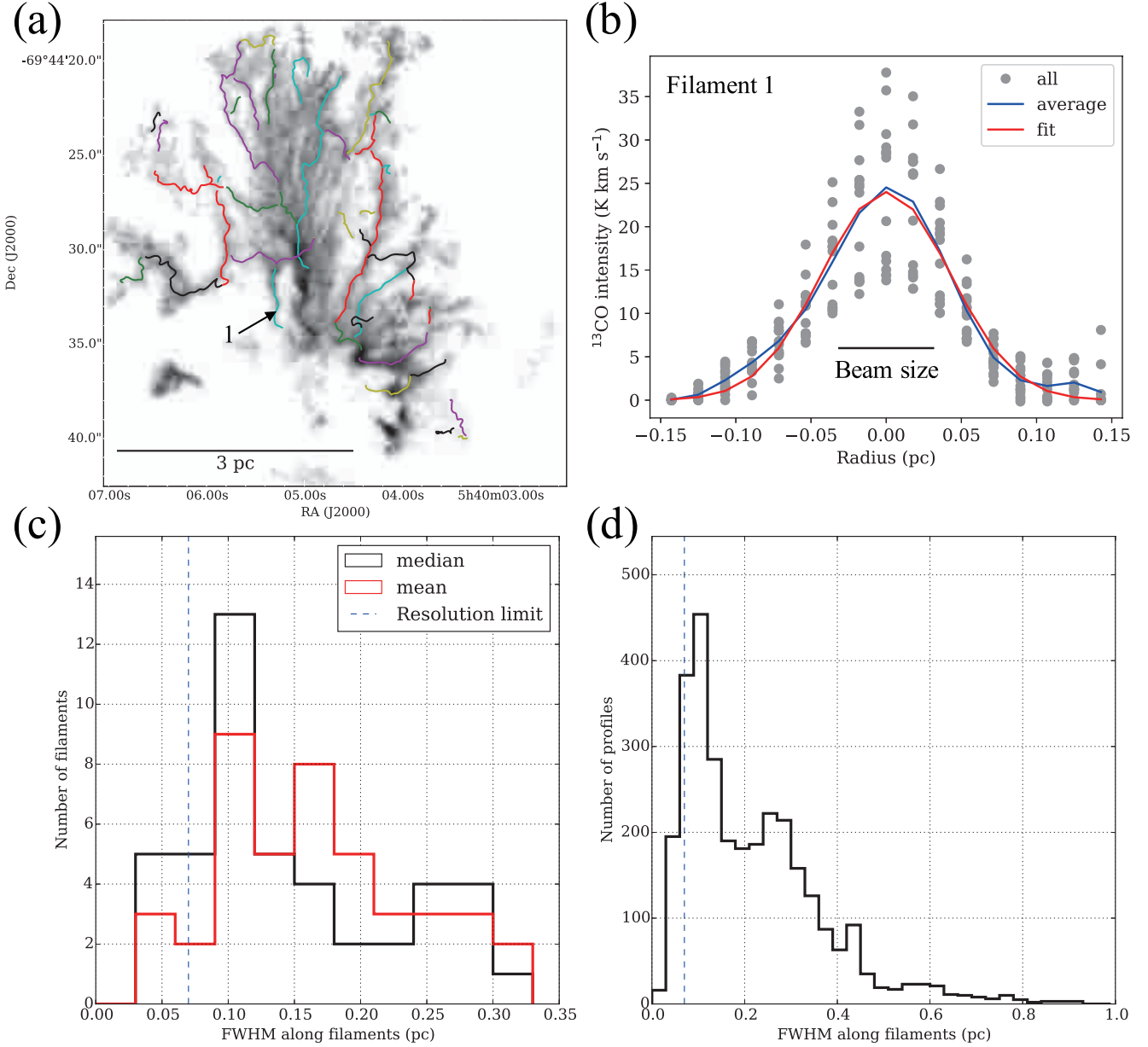}
\caption{Filamentary structures and their width toward the N159E-Papillon region. (a) Color lines show the identified filaments (see the text) on the $^{13}$CO velocity-integrated intensity map. {(b) The intensity profile of filament 1. The grey dots are all the points of the cross sections taken along the filament. The blue and red lines show the averaged profile and the Gaussian fitting result for the profile, respectively. } (c) Distributions of the FWHM widths for identified filaments. {(d) The number of profiles (every radial slice) along all the identified filaments.}\label{fig:Dis}}
\end{figure}

The previous observations in the Cycle 1 with an angular resolution of $\sim$1$\arcsec$ ($\sim$0.24\,pc) did not resolve the thin filaments; however, the Cycle 4 data clearly identified much narrower filaments that were previously unresolved. 
To identify the filamentary structures from the $^{13}$CO data quantitatively, we applied the DisPerSE algorithm \citep{Sousbie11}, which has been developed to extract structures such as filaments and voids in astrophysical data. This algorithm was applied to the dust continuum data obtained by the {\it Herschel} Gould Belt survey \citep[e.g.,][]{Andre10,Doris11} and molecular line emission in Taurus \citep{Panopoulou14}. {To extract topological components from the molecular line cube, we} set a $``$persistence$"$ threshold, and the difference in value between the two points in the pair lower than the given threshold is cancelled \citep[see][]{Sousbie11}, of 20 K in brightness temperature of $^{13}$CO\,($J$ = 2--1), which roughly corresponds to the typical peak temperature in the high-column density ($\gtrsim$5\,$\times$\,10$^{22}$ cm$^{-2}$) regions. 
{We defined structures with an aspect ratio $>$3:1 as the filaments after the measurements of the width (see below) and length. For this reason, some of the pillars (Sect. \ref{R:Pillar}) and high-column density regions around the Papillon Nebula are not identified in this criterion. In total,} we identified $\sim$50 filaments that are overlaid on the $^{13}$CO velocity-integrated  intensity map in Figure \ref{fig:Dis} (a). To estimate filamentary widths, we performed Gaussian fits to the normalized radial intensity profile to the filament along the identified skeleton shown in Figure \ref{fig:Dis} (a). {An example of the fitted profile is shown in the panel (b).}
The resultant median FWHM widths along the filaments ($W_{\rm fil}$), shown in Figure \ref{fig:Dis} (b), has a remarkable peak at $\sim$0.1\,pc with the standard deviation of 0.08\,pc.
The widths of most of the filaments are significant at the present resolution ($\sim$0.07\,pc). Furthermore, the derived width is not an effect of the averaging of the radial profile across all the identified filaments, as shown in Figure \ref{fig:Dis} (d). 
Such a small filamentary width is similar to the Galactic star-forming regions reported by \cite{Doris11,Doris19} and others, and is the first detection of such narrow filaments outside our Galaxy (see also THF18).
We estimated the column density and mass of the filamentary clouds using the $^{12}$CO and $^{13}$CO data by assuming the local thermodynamical equilibrium. We obtained the excitation temperature ($T_{\rm ex}$) map from the $^{12}$CO data assuming that the line is optically thick. Then we got the optical depth of $^{13}$CO using the $T_{\rm ex}$ map, and finally, the $^{13}$CO column density was derived using the previous calculations. We adopted a $^{13}$CO/H$_2$ abundance ratio of 3.1\,$\times$\,10$^{-7}$, taking $^{12}$CO/H$_2$ as 1.6\,$\times$\,10$^{-5}$ and $^{12}$CO/$^{13}$CO as 50 (see also \cite{Mizuno10} and the references therein), to obtain the H$_2$ column density. 
The cumulated mass of all structures shown in Figure \ref{fig:13COfilament} using the mean molecular weight, $\mu$, of \citep{Kauffmann08} is $\sim$1\,$\times$\,10$^4$ $M_{\odot}$, which corresponds to $\sim$10--20\% of that of the entire molecular cloud in the N159E region (Paper I). The averaged column density of {the N159E-Papillon region shown in Figure \ref{fig:13COfilament}} is calculated to be $\sim$5\,$\times$\,10$^{22}$\,cm$^{-2}$. The central column densities at the filament crests ($\Sigma_{\rm 0}$) are $\sim$(5--30)\,$\times$\,10$^{22}$\,cm$^{-2}$ and the resultant line masses ($M_{\rm line}$) are calculated to be $\sim$(1--7)\,$\times$\,10$^2$\,$M_{\odot}$\,pc$^{-1}$ using the equation $M_{\rm line} \approx \Sigma_{\rm 0} \times {\rm W_{\rm fil}}$ \citep[e.g.,][]{Andre14}. 
This is consistent with filaments found in the Galactic high-mass star-forming regions (\citealt{Hill12} for Vela C; \citealt{Andre16} for NGC\,6334). Such filaments with very large line-mass are supposed to be quite unstable against the gravitational collapse unless we assume that there is an additional support force, such as strong magnetic fields (see also discussions in \citealt{Andre16}).

\subsection{CO Pillars around the Papillon Nebula \label{R:Pillar}}
\begin{figure}[htbp]
\includegraphics[width=180mm]{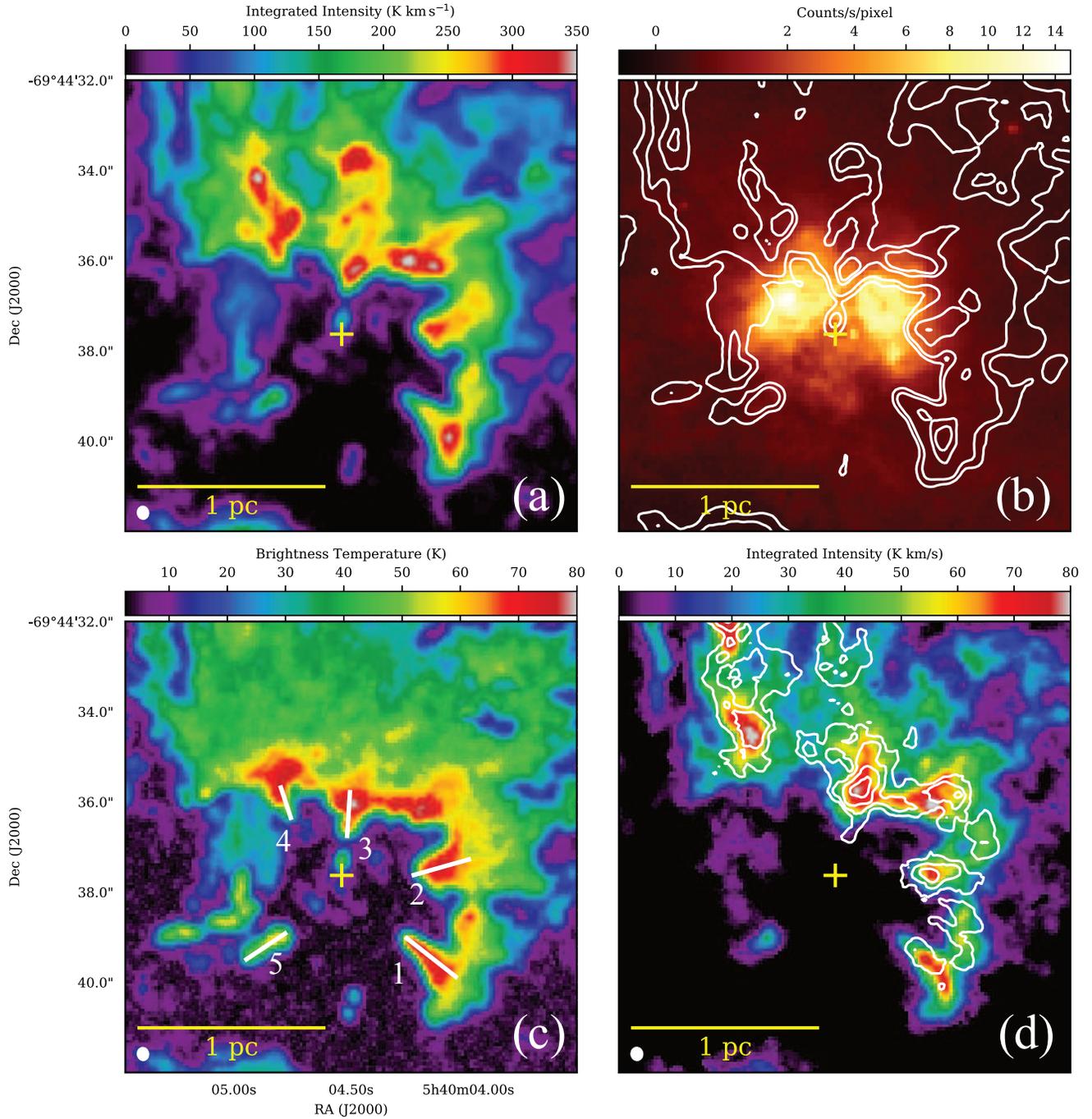}
\caption{
Zoomed-in views of CO, its isotopes, and H$\alpha$ emission in the Papillon Nebula. (a) Total velocity-integrated intensities of $^{12}$CO\,($J$ =  2--1) are shown in color scale. (b) Color scale and contours of the H$\alpha$ emission obtained with {\it HST} and the $^{12}$CO integrated intensity as in panel (a), respectively. The CO contour levels are 50, 100, 250, and 300\,K\,km\,s$^{-1}$. (c) Color-scale image of the peak brightness temperature map of $^{12}$CO\,($J$ = 2--1). (d) Color-scale image and contours of total velocity-integrated intensities of $^{13}$CO\,($J$ = 2--1) and C$^{18}$O\,($J$ = 2--1), respectively. The C$^{18}$O contour levels are 0.5, 1.5, 2.5, and 3.5\,K\,km\,s$^{-1}$. The yellow crosses in all panels represents the position of the Papillon Nebula YSO. The angular resolutions of the ALMA data are shown in the lower left corners in each panel.
\label{fig:COpillars}}
\end{figure}

\ Figure \ref{fig:COpillars} shows the $^{12}$CO (panels (a)--(c)), $^{13}$CO, C$^{18}$O\,($J$ = 2--1) (panel (d)) and the H$\alpha$ (panel (b)) distributions around the Papillon Nebula. Our previous observations confirmed that there is a CO cavity {centered on the Papillon Nebula YSO} (Paper I). 
{The present high-resolution data clearly resolved several elongated structures along the H${\alpha}$ bright region, the Papillon Nebula, as shown in Figure \ref{fig:COpillars} (b). 
These structures roughly extend toward the position of the Papillon Nebula YSO and are considered to be heated by the protostars and the H$\;${\sc ii} region because they show high-brightness temperature (60--80\,K) in $^{12}$CO (Figure \ref{fig:COpillars} (c)). We define these structures as $``$CO pillars$"$ by eye based on the abovementioned features, independently of the filament identification (Sect. \ref{R:13COfilament}), and the CO pillars are shown by white lines in Figure \ref{fig:COpillars} (c).}
This is the first detection of $``$pillars of creation$"$ in an external galaxy (see \citealt{Hester96} for the Eagle Nebula). 
The properties of the CO pillars are listed in Table \ref{table:Pillar}.
The averaged column densities of the CO pillars are (0.9--3.0)\,$\times$\,10$^{23}$\,cm$^{-2}$.  
We derived the width (FWHM) of the pillars by Gaussian fitting to the averaged column density profile perpendicular to the white lines in Figure \ref{fig:COpillars}. The resultant width of the pillars is {$\sim$0.1--0.2\,pc}, which is similar to that of the filaments in the region. The average volume density, $n$(H$_2$), is $\sim$(1--7)\,$\times$\,10$^5$\,cm$^{-3}$. 
These results show that the density and temperature are significantly higher than those of the surrounding filaments. The high-density nature is also supported by the detection of the C$^{18}$O\,($J$ = 2--1) emission along the CO pillars (Figure \ref{fig:COpillars} (d)). 
We do not see clear extinctions of the H$\alpha$ emission along the CO pillars, except for the central part of the Papillon Nebula, despite its high-column density. This indicates that the pillars are embedded in the ionized region or located behind it. 

\begin{deluxetable}
{ccccccc}  
\tabletypesize{\scriptsize}
\tablecaption{{Properties of CO pillars around the Papillon Nebula} \label{table:Pillar}}
\tablewidth{0pt}
\tablehead{
Name &$T_{\rm max}^{^{12}\rm CO}$ (K)$^{\rm a}$ & $N({\rm H_2})_{\rm max}$ (cm$^{-2}$)$^{\rm b}$ & $N({\rm H_2})_{\rm ave}$ (cm$^{-2}$)$^{\rm c}$ & Width (pc)$^{\rm d}$ & $n({\rm H_2})_{\rm ave}$ (cm$^{-3}$)$^{\rm e}$ }
\startdata
Pillar 1 & 77 & 4.0\,$\times$\,10$^{23}$ & 2.5\,$\times$\,10$^{23}$ & 0.12 & 6.7\,$\times$\,10$^{5}$ \\
Pillar 2 & 76 & 3.7\,$\times$\,10$^{23}$ & 2.6\,$\times$\,10$^{23}$ & 0.18 & 4.7\,$\times$\,10$^{5}$ \\
Pillar 3 & 80 & 4.1\,$\times$\,10$^{23}$ & 3.0\,$\times$\,10$^{23}$ & 0.22 & 4.4\,$\times$\,10$^{5}$ \\
Pillar 4 & 67 & 1.3\,$\times$\,10$^{23}$ & 1.0\,$\times$\,10$^{23}$ & 0.13 & 2.7\,$\times$\,10$^{5}$ \\
Pillar 5 & 61 & 1.4\,$\times$\,10$^{23}$ & 9.0\,$\times$\,10$^{22}$ & 0.09 & 3.2\,$\times$\,10$^{5}$ \\
\enddata
\ \ {\tablenotetext{\rm a}{Maximum brightness temperature of $^{12}$CO\,($J$ = 2--1).} 
\tablenotetext{\rm b}{Maximum H$_{2}$ column density.}
\tablenotetext{\rm c}{Averaged H$_{2}$ column density within the certain regions. In order to separate from the surrounding structures, the boundaries of Pillar 1, 2, 3, and 5 were defined as 50\%, 60\%, 70\%, 50\%, respectively, of the maximum column densities of each pillar. The boundary of Pillar 4 is manually defined at the edge of the white line because we cannot divide the surrounding high-column density gas located at the northern part of the pillar with the similar criterion.} 
\tablenotetext{\rm d}{FWHM of the pillars (see the text).}
\tablenotetext{\rm e}{Averaged H$_{2}$ volume density obtained by dividing the $N_{\rm ave}$ by the width.}}
\end{deluxetable}

\subsection{Protostellar sources with molecular outflow \label{R:outflow}}
\begin{figure}[htbp]
\includegraphics[width=180mm]{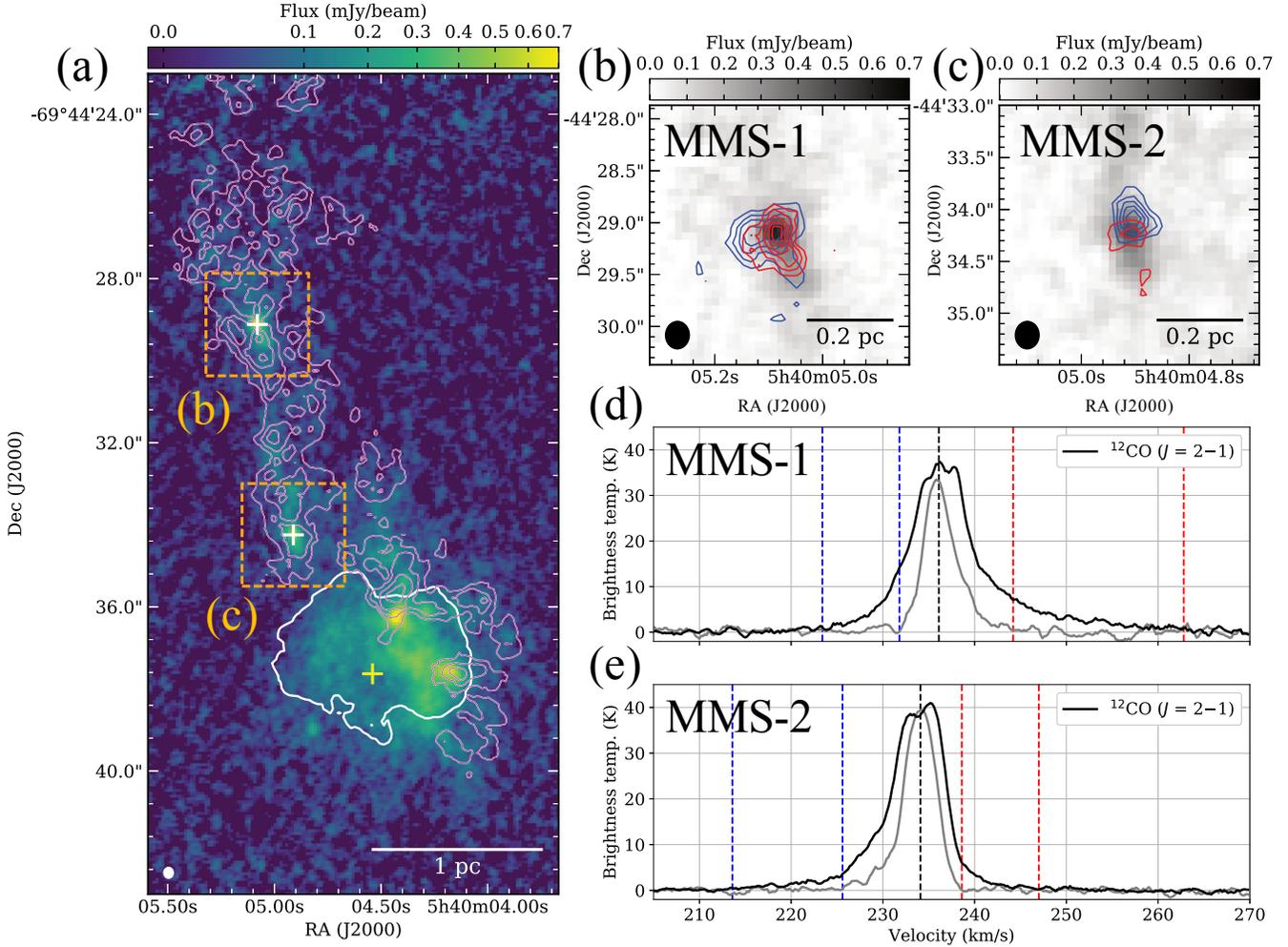}
\caption{
Distributions of high-velocity outflow wings and filamentary structures traced by continuum/line emissions toward the N159E-Papillon region. (a) Color-scale image and lilac contours of 1.3\,mm continuum and velocity-integrated intensities of C$^{18}$O\,($J$ = 2--1) emission, respectively. The contour levels are 0.5, 1.5, 2.5, and 3.5\,K\,km\,s$^{-1}$. The white/yellow crosses and the white contour are the same as those in Figure \ref{fig:13COfilament}. (b) and (c) Enlarged views of 1.3\,mm continuum and outflow wings toward MMS-1 and 2. The grayscale images show the 1.3\,mm continuum. The blue and red contours show images of $^{12}$CO\,($J$ = 2--1) integrated over the velocity ranges shown by red and blue dashed lines in panels (d) and (e). 
The lowest contour and the subsequent contour step are 10 and 5\,K\,km\,s$^{-1}$, respectively.
(d) and (e) Black lines show the averaged intensity profiles of $^{12}$CO\,($J$ = 2--1) toward the 1.3\,mm continuum peaks of MMS-1 and MMS-2. The gray lines show the reference $^{12}$CO spectra at the positions of $\sim$1$\arcsec$ away from the millimeter sources. The black dotted lines show the systemic velocities estimated from Gaussian fitting to the $^{13}$CO profile at each continuum peak.
\label{fig:outflow}}
\end{figure}

Figure \ref{fig:outflow} (a) shows the 1.3\,mm continuum and C$^{18}$O\,($J$ = 2--1) distribution around the Papillon Nebula. 
The extended 1.3\,mm continuum emission around the Papillon Nebula YSO roughly agrees with the distribution of the H${\alpha}$ emission and has no counterparts of molecular gas, indicating that the continuum emission seems to be dominated by free-free emission from the ionized gas (see also Paper I).
On the other hand, the filamentary dust clumps along the C$^{18}$O emission, a tracer of cold/dense molecular gas, are considered to be dominated by the thermal dust emission.
Two major local maxima, hereafter MMS-1 and MMS-2, are found in the 1.3\,mm continuum image in Figures \ref{fig:outflow} (a-c). Because we detected extended emission in 1.3\,mm and C$^{18}$O along the north--south direction, the two sources are considered to be dense cores embedded in the filamentary cloud. The peak column densities and total masses of both sources deduced from the 1.3\,mm continuum emission are $\sim$1\,$\times$\,10$^{24}$ cm$^{-2}$, and $\sim$2\,$\times$\,10$^2$ $M_{\odot}$, respectively, with an assumption of the dust emissivity, $\kappa_{\rm 1.3 mm}$ of 1\,cm$^{2}$\,g$^{-1}$ for protostellar envelopes \citep[e.g.,][]{Oss94}, a dust-to-gas ratio of 3.0\,$\times$\,10$^{-3}$ \citep{Herrera13,Gordon14} and a uniform dust temperature of 20 K, which is typically adapted/estimated for galactic high-mass star-forming dust clumps \citep[e.g.,][]{Urquhart14,Yuan17}. Note that the core boundaries deriving their total masses were set as the 30\% intensity level of each continuum peak to avoid the contaminations from the parental filamentary clouds. Although these sources were not cataloged as point sources in the infrared observations so far \citep[e.g.,][]{Chen10,Carlson12}, their detection in the 1.3\,mm is strongly suggestive of their protostellar nature. We have identified outflow wings that have a velocity span of $\sim$30 km\,s$^{-1}$ in $^{12}$CO\,($J$= 2--1) toward MMS-1 and MMS-2. Figures \ref{fig:outflow} (b-e) show the distributions and the line profiles of the outflow wings. 
We extracted the spectra and chose the velocity ranges using the following procedure. We defined the outer boundaries of the outflow spectra (i.e., maximum relative velocity) above the 2$\sigma$ level of the velocity-smoothed spectra with a smoothing kernel of 9\,ch (=1.8\,km\,s$^{-1}$). 
We extracted the spectra at positions $\sim$1$\arcsec$ away from the millimeter sources in the east direction as the references without the outflow contaminations and defined the velocities below 2$\sigma$ intensity levels as the inner edge of the outflow spectra (see Figure \ref{fig:outflow}(d) and (e)). The outflow and YSO properties are listed in Table \ref{table:Outflow}. The size of the outflow is as small as $\sim$0.1\,pc, and escaped detection with our lower-resolution study (Paper I). It is likely that the launching points of the outflows coincide with the 1.3\,mm continuum peaks. Because the positional offsets between the blue-shifted and red-shifted lobes are quite small toward MMS-1 and MMS-2, the outflow orientations may be close to pole-on or very small. The dynamical time ($t_{\rm d}$) of the outflows is roughly estimated to be $<$10$^4$\,yr from a ratio of 0.1\,pc/20 km\,s$^{-1}$. We roughly estimated the mechanical forces of the outflow lobes ($F_{\rm out}$) using the following equation: $F_{\rm out}$ = $M_{\rm out}V_{\rm out}$/$t_{\rm d}$ \citep[e.g.,][]{Beuther02}, where $V_{\rm out}$ and $M_{\rm out}$ are the outflow velocity and mass, respectively. 
The estimated $F_{\rm out}$ is {$\sim$10$^{-3}$--10$^{-2}$\,$M_{\odot}$\,km\,s$^{-1}$\,yr$^{-1}$, depending on the assumption of the outflow inclination angle ($i$ = 30$\arcdeg$--70$\arcdeg$). The $F_{\rm out}$ and the envelope mass derived from millimeter dust continuum observations} are consistent with those of Galactic high-mass protostars \citep[e.g.,][]{Beuther02} and the N159W-South region (TFH19). 
A large amount of the surrounding gas with a mass of $\sim$10$^{2}$\,$M_{\odot}$ and the nondetection of infrared emission suggest that MMS-1 and MMS-2 are in an extremely young phase of high-mass star formation, with an age of $\lesssim$10$^{4}$\,yr. 
We note that the Papillon Nebula YSO is more evolved than the two sources; however, the age is as young as $\sim$0.1\,Myr (Paper I) judging from the combination of the spectral energy distribution (SED) fitting using the {\it Spitzer/Herschel} data and the compactness of the H$\;${\sc i} region (see also \citealt{Chen10}). 
Although certain time differences are found, the three high-mass protostellar systems with separations of $\sim$1--2\,pc are formed within the order of $\sim$0.1\,Myr.
We discuss the possible processes of high-mass star formation taking place in the N159E-Papillon region in Sect. \ref{D:filament}.

\begin{deluxetable}
{ccccccc}  
\tabletypesize{\scriptsize}
\tablecaption{YSO characteristics and outflow properties in the N159E-Papillon region \label{table:Outflow}}
\tablewidth{0pt}
\tablehead{
\multirow{2}{*}{Source name} &\multirow{2}{*}{$M^{*}$ ($M_{\odot}$)$^{\rm a}$} & \multirow{2}{*}{$L^{*}$ ($L_{\odot}$)$^{\rm b}$} & $M_{\rm out}$ ($M_{\odot})^{\rm c}$ & Distance (pc)$^{\rm d}$ & Velocity (km\,s$^{-1}$)$^{\rm e}$ \\
		                              &		                                                    &		                      &			Blue/Red			        & 			Blue/Red	& 		Blue/Red }
\startdata
Papillon Nebula YSO   & 21 + 41  & (0.5--1.8)\,$\times$\,10$^{5}$ & $\cdots$ & $\cdots$ & $\cdots$ \\
MMS-1 & $\cdots$   & $\cdots$ & 4.8/3.4 & $\lesssim$0.1 & {12.7/26.7} \\
MMS-2 & $\cdots$   & $\cdots$ & 2.5/1.3 & $\lesssim$0.1 & {20.5/12.9} \\
\enddata
\tablenotetext{\rm a} {References: \cite{Heydari99,Testor07} for the 21\,$M_{\odot}$ protostar, and \cite{Indebetouw04,Chen10} for the 41\,$M_{\odot}$ one.}
\tablenotetext{\rm b}{The average total luminosity; see \cite{Carlson12} and \cite{Chen10} for details.}
\tablenotetext{\rm c}{The outflow mass is estimated from the $^{12}$CO\,($J$ = 2--1) intensity by assuming a conversion factor form $^{12}$CO\,($J$ =1--0) intensity to the column density of $X_{\rm co}$ = 7\,$\times$\,10$^{20}$ cm$^{-2}$ (K\,km\,s$^{-1}$)$^{-1}$ cm$^{-2}$ \citep{Fukui08} and the $^{12}$CO\,($J$ = 2--1)/$^{12}$CO\,($J$ = 1--0) ratio = 1.0.} 
\tablenotetext{\rm d}{Projected distances to the peak intensity of the outflow lobes from the continuum peaks.}
\tablenotetext{\rm e}{Maximum radial velocity of the outflow lobe with respect to the systemic velocity (see also the black lines in Figures \ref{fig:outflow}(d) and (e)).}
\end{deluxetable}

\section{Discussions \label{Dis}} 

\subsection{Filament/Protostar formation in the N159E-Papillon region \label{D:filament}}
\begin{figure}[htbp]
\includegraphics[width=180mm]{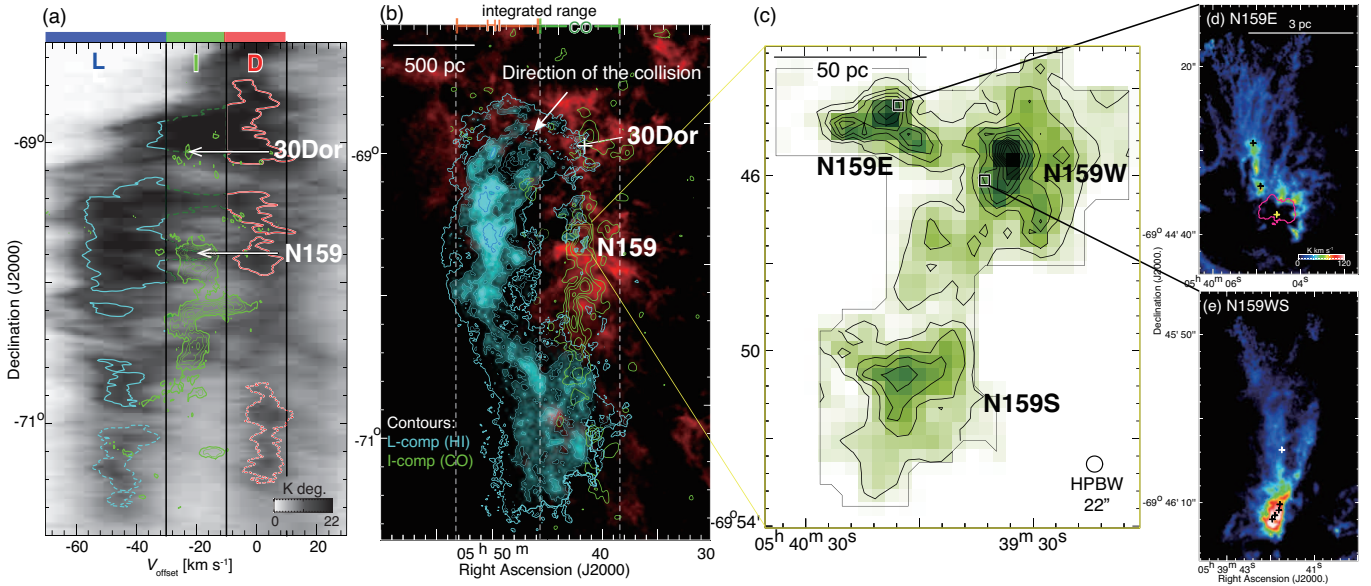}
\caption{
Large- and small-scale distributions of H$\;${\sc i} and CO emission around the N159 GMC. (a) The position-velocity diagrams of H$\;${\sc i} \citep{Kim03} and CO\,($J$ = 1--0) \citep{Fukui08} are shown in grayscale and green contours, respectively. The extracted ranges to make the maps are shown as white dashed lines in panel (b). (b) Cyan and red images of H$\;${\sc i} integrated intensity maps of L- ($V_{\rm offset}$ = $-$100 to $-$30\,km\,s$^{-1}$), and the D-component ($V_{\rm offset}$ = $-$10--10\,km\,s$^{-1}$), respectively, reproduced from \cite{Fukui17}, and \cite{Tsuge19}. The green contours show the integrated intensity map of CO\,($J$ = 1--0) with an angular resolution of 2\farcm7. (c) The green image and contours show the CO\,($J$ = 3--2) map of the N159 GMC \citep{Minamidani08}. (d) and (e) ALMA $^{13}$CO images of the N159E-Papillon (same as Figure \ref{fig:13COfilament} (a)) and the N159W-South regions (TFH19). The shape of the Papillon Nebula traced by H$\alpha$ emission is shown in magenta contours in panel (d). The positions of the protostellar sources, the Papillon Nebula YSO, and outflow sources with and without 1.3\,mm continuum sources are denoted by the yellow, black, and white crosses, respectively.
}
\label{fig:preHICO}
\end{figure}

We discuss the formation of the protostars and the filaments in the N159E-Papillon region. In Paper I, we discussed the possibility that collisions by three molecular filaments induced the star formation activity in the Papillon Nebula. However, such a filamentary cloud collision model may not be appropriate to explain the star formation activities in this region revealed by the present observations for some reasons. (1) The present observations clearly show that the parental molecular cloud is composed of many complex filamentary structures (Figures \ref{fig:13COfilament} (a) and \ref{fig:Dis}) rather than the previously identified several filaments (Figure \ref{fig:13COfilament} (b)). Therefore, it is more likely that the molecular cloud was formed in an entangled filamentary configuration rather than by a coalescence process of many individual filaments. (2) The previous lower-resolution observations could not find the protostellar activities in MMS-1/2 (Sect. \ref{R:outflow}), and thus it was thought that the high-mass star-formation seen at the Papillon Nebula is a local event around this region. Despite this, the projected separations between the Papillon Nebula YSO and MMS-1/2 are as large as $\sim$1--2\,pc, and these objects are independently formed within a few 0.1\,Myr.
In addition to this, our separate observations toward the N159W-South region (TFH19), which is located $\sim$50\,pc apart from the N159E-Papillon cloud in projection, found that both regions show quite similar star-formation activities and molecular gas distributions (see also Figure \ref{fig:preHICO}).  In N159W-South, we detected multiple protostellar sources with a separation of $\sim$1--2\,pc, along with the massive molecular filaments (TFH19). The physical properties of the outflows and molecular filaments are quite similar to those in the N159E-Papillon region and the global orientation of both filaments is roughly parallel to each other. These facts indicate that the star formation activity and the filament formation taking place in the N159 region are not only local phenomena with a scale of 1--2\,pc, but also a more global event extending up to $>$50\,pc. We thus consider the large-scale gas kinematics as an origin of the formation of massive protostars/filaments in this region.

We calculated the filament width to be $\sim$0.1\,pc, which is reported in a number of Galactic studies (e.g., \citealt{Andre14} and references therein) and its high line mass, $\sim$a few\,$\times$\,100 $M_{\odot}$\,pc$^{-1}$. Such high line-mass filaments are unstable against the gravitational collapse unless some additional strong support forces emerge to regulate and slow down the collapse. There is a possibility that the massive filaments were formed by a large-scale converging flow quite recently with a timescale of $\sim$a few 0.1\,Myr (see also Sect. \ref{sec:intro}). If we ignore the magnetic field, the virial mass per unit length, $M_{\rm line,vir}$ = 2$\sigma_{\rm v}^{2}$/G \citep[see][]{Fiege00}, is calculated to be $\sim$1.4\,$\times$\,10$^2$\,$M_{\odot}$\,pc$^{-1}$, as the typical velocity dispersion, $\sigma_{\rm v}$, is $\sim$0.6\,km\,s$^{-1}$ measured from the present $^{13}$CO data. The $M_{\rm line,vir}$ is smaller than the observed line mass (Sect. \ref{R:13COfilament}) and it is likely that the filaments are gravitationally unstable. \cite{Nayak18} also reported the low virial parameters in molecular clumps in the N159 region, thus we suggest that the filaments may not be long-lived objects. Even if the strong magnetic field of $\sim$300\,$\mu$G is penetrated perpendicular to a $\sim$0.1\,pc width filament, a line mass of $\sim$70\,$M_{\odot}$\,pc$^{-1}$ is a limit that can be magnetically supported according to \cite{Inoue18}.

\begin{figure}[htbp]
\includegraphics[width=180mm]{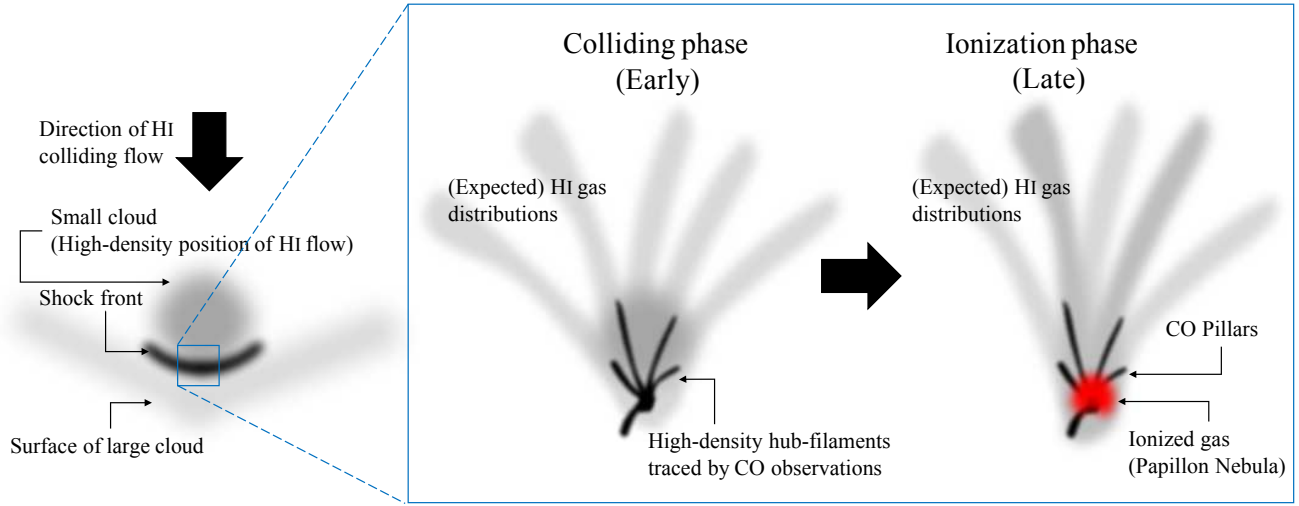}
\caption{
Schematic views of the collisional event in the N159E-Papillon region. {A large-scale view of the colliding event with the size of more than 10\,pc is shown in the left panel. The zoom on views in the middle and right panels show what is happening around the shock front. Black and gray colors represent the CO distributions traced by the ALMA observations and expected H$\;${\sc i} gas distributions, respectively.} 
\label{fig:Ponti}}
\end{figure}

We summarize the early atomic/molecular gas studies around the N159 region. Figure \ref{fig:preHICO} shows the large-scale gas distributions seen in H$\;${\sc i} \citep{Kim03} and CO (\citealt{Fukui08,Minamidani08}, see also \citealt{Wong11}). \cite{Fukui17} subtracted the galactic rotation from the H$\;${\sc i} data and found three major velocity features, L-, I-, and D-components \citep[see also][]{Luks92,Tsuge19}. 
The largest giant molecular complex in the LMC, called $``$the molecular ridge,$"$ which is shown in green contours in Figures \ref{fig:preHICO}(a) and (b), is mainly located at the intermediate velocity range between the L- and D-components \citep{Fukui17}. These authors proposed that the tidal interaction between the LMC and the Small Magellanic Cloud (SMC) drove a kpc H$\;${\sc i} flow, which originated in the past close encounter between the two galaxies, which was modeled in the numerical simulations by \cite{Bekki07} following the original suggestion by \cite{Fujimoto90}. After {settling} outside the LMC disk for 200\,Myr, the gas could be now inducing the formation of the super star cluster R136 in 30 Dor as well as the molecular ridge containing the N159 region {\citep{Fukui17}}. Because the GMC in the N159 region is the most massive molecular reservoir of the region \citep[e.g.,][]{Fukui08,Minamidani08,Chen10}, {this GMC may be the end product of a cloud strongly compressed by the H$\;${\sc i} flow and enriched to the large mass we observe currently \citep{Fukui17}}. Recent MHD simulations by \cite{Inoue18} investigated detailed processes of filament/protostar formation driven by cloud--cloud collisions. These authors demonstrated that massive hub filaments with their complex morphology that qualitatively resembles the filaments in the N159E-Papillon region, are formed in the compressed layer between the cloud and the flow, within a few\,$\times$\,0.1\,Myr after the collisional event. 
The colliding flow or GMC is likely moving from the northwest to the southeast, as suggested by \cite{Fukui17} (see also the white arrow in Figure \ref{fig:preHICO} (b)), and the global orientation of the molecular filaments in the N159E/W region roughly follows the flow (Figures \ref{fig:preHICO}(d) and (e)). These features are reproduced by numerical simulations of colliding clouds, as demonstrated by \cite{Inoue18}. The present case in the N159E-Papillon region shows that the typical lengths of the filaments are $\sim$6\,pc with a vertical angle of $\sim$90$\arcdeg$ and they are in a north--south projected direction, as schematically shown in Figure \ref{fig:Ponti} (see also Figure \ref{fig:Dis} (a)). This distribution can be explained as follows: the initial small cloud collided with the extended cloud as simulated by \cite{Inoue18}. 
As shown in Figures \ref{fig:preHICO}(a) and (b), the distribution of H$\;${\sc i} gas is inhomogeneous, and thus the colliding flow is considered to have local high-density parts. Such a kind of colliding flow can mimic a collision between a small cloud and a large one (Figure \ref{fig:Ponti}).

The observed filamentary cloud has a velocity gradient of $\sim$1.5\,km\,s$^{-1}$\,pc$^{-1}$ in the north--south direction (Figures \ref{fig:13COchanmap} and \ref{fig:Moment1}(a) and (b)). Given the filament formation scenario based on the H$\;${\sc i} colliding flow (Figures \ref{fig:preHICO} and \ref{fig:Ponti}, \citealt{Fukui17}), we have a clear picture of the gas flow in the N159 region. Assuming that the clump interacting with the D-component is falling from the northwest (Figure \ref{fig:Ponti}), it is likely that the clump comprises the hub of the N159E filamentary system. We expect that the filaments are decelerated via the frictional interaction with the ambient H$\;${\sc i} gas more than the dense hub. This deceleration will cause a velocity gradient, as the hub is moving toward us, and is more blueshifted than the less dense filaments, which explains an explanation on the velocity gradient. The N159W-South also has a similar interaction with the disk H$\;${\sc i}, and shows a similar velocity gradient in the same sense (TFH19), although the observed gradient is smaller than that of the N159E filamentary system, possibly due to the projection effect. It is thus a natural outcome that the two filamentary systems have a common velocity gradient in spite of a separation of $\sim$50\,pc.
 
Assuming that the ionized shock front proceeds at 5\,km\,s$^{-1}$ (e.g., \citealt{Fukui16} for RCW38), the dynamical time scale of the interaction is $\sim$1.2\,Myr, as the size (length) of the molecular clouds is $\sim$6\,pc. The interaction will result in the formation of an O-type star and the resulting ionization of a cloud in the following 0.1\,Myrs. This means that the lower limit of the CO filament lifetime after their formation is of the order of $\sim$1\,Myr, which is consistent with our argument that the CO filaments were formed very recently, as discussed in the previous paragraphs.
The collision took place between two velocities of H$\;${\sc i} flow \citep{Fukui17}, whereas the currently available H$\;${\sc i} data with an angular resolution of $\sim$1\arcmin\,\citep{Kim03} are not high enough to distinguish the relation between the molecular filaments and the large-scale H$\;${\sc i} gas.
The initial H$\;${\sc i} gas still needs to be observed at a resolution comparable to that of our ALMA observations to better understand its connection with the molecular reservoirs.

\subsection{Comparison with Galactic Pillars of creation \label{D:pillar}}
\ It is also important to better understand how feedback is affecting the environment of high-mass stars in disentangling the gas distribution that is usually a mixture of the initial conditions prior to high-mass star formation and the results of feedback effects. In that context, we consider the formation mechanism and the evolutionary state of the CO pillars around the Papillon Nebula. In this section, we compare our study with the early results for the Eagle Nebula in M16, the most famous $``$pillars of creation$"$ in the Galaxy, and compare them with the present results in N159E-Papillon. \cite{Pound98} observed the Eagle Nebula in $^{12}$CO, $^{13}$CO, and C$^{18}$O\,($J$ = 1--0) by using the Berkeley--Illinois--Maryland Array with a resolution of $\sim$0.1\,pc. The mean column density and volume density deduced from the CO observations are 2\,$\times$\,10$^{22}$ cm$^{-2}$ and $\sim$3\,$\times$\,10$^3$ cm$^{-3}$, respectively. The total ionizing flux is $S\sim$1.2\,$\times$\,10$^{50}$ s$^{-1}$, mostly from two O5 stars in M16 and the separation between the ionizing stars and the pillars is a few parsecs \citep{Hillenbrand93}. The age of M16 is estimated to be 1.3$\pm$0.3\,Myr \citep{Bonatto06}. The Papillon Nebula, with an age of $\sim$0.1\,Myr (Paper I) is in a much younger stage than the Eagle Nebula and the separation of the pillars is $\sim$0.5\,pc from the YSO. Although the total ionization fluxes of both regions are almost the same or higher in the Papillon Nebula ($S$ $\gg$ 1.1\,$\times$\,10$^{50}$ s$^{-1}$, {Paper I}), the column density and volume density of the CO pillars in the Papillon Nebula, {(0.9--3.0)}\,$\times$\,10$^{23}$\,cm$^{-2}$ and $\sim${(3--7)}\,$\times$\,10$^{5}$\,cm$^{-3}$, respectively, are an order of magnitude higher than those of the Eagle Nebula. This indicates that dissociation of the molecular gas toward the CO pillars has not reached the innermost part. On the other hand, because the boundaries of the $^{12}$CO, $^{13}$CO, and C$^{18}$O distributions facing the H$\alpha$ emission are very close to each other (Figure \ref{fig:COpillars}), the surrounding low-density envelopes of the CO pillars have already been dissociated.

We investigate the velocity structure of the CO pillars to explore the formation mechanism. If the Rayleigh-Taylor instability \citep{Spitzer54} is working to develop pillar-like objects, the velocity gradient should increase with the square root of the distance from the top of the pillar \citep{Frieman54}. \cite{Pound98} pointed out the Rayleigh-Taylor instability is unlikely, as the formation process of the pillars in the Eagle Nebula is based on the CO velocity fields. Our observations also show similar velocity gradients to what is shown in Figure 5 of \cite{Pound98} or no significant velocity gradient, as shown in Figure \ref{fig:Moment1}. We thus exclude the Rayleigh-Taylor instability as a candidate mechanism to form the CO pillars in the Papillon Nebula. Alternatively, we suggest that the radially distributed hub-filaments were the initial configuration before the high-mass star formation (see also Figure \ref{fig:Ponti}).  The present gas distribution of the Papillon Nebula (Figure \ref{fig:COpillars}) can be explained as follows: the high-mass star was formed at the intersection of the hub and the central part and the relatively low-density gases around the filaments were dissociated. Hub filaments without any ionization regions are also seen in the N159W-South clump, where ionization has not yet started (Paper I; TFH19) and it may be in an earlier stage than the Papillon Nebula. N159W-South may possibly evolve later into a stare similar to the N159E-Papillon when the protostar becomes a mature O star. 

\section{Conclusions \label{sec:conclusions}}
 We carried out high-resolution observations of the N159 E region in the ALMA Cycle 4 period. Our main conclusions can be summarized belows:
\begin{enumerate}
\item The cloud distribution was known to be filamentary at $\sim$1\arcsec resolution based on the Cycle 1 observations (Paper I). The present observations revealed that the previously identified filamentary distribution consists of even smaller filaments with a typical width of 0.1\,pc at $\sim$0\farcs28 resolution in the $^{13}$CO\,($J$ = 2--1) transition. These CO filaments appear to be extended nearly radially over 1--2\,pc, with a pivot toward the Papillon Nebula, which is ionized by an O star. The line mass of the filaments is estimated to be $\sim$(1--7)\,$\times$\,10$^2$ $M_{\odot}$\,pc$^{-1}$ and is comparable to that in high-mass star-forming regions in the Milky Way.
\item One of the filaments is associated with the other two protostars driving molecular outflows with a dynamical timescale of $\sim$10$^4$ yr, which is presumably two high-mass protostars located at 1--2\,pc north of the Papillon Nebula. This indicates very recent formation of at least three high-mass protostellar systems separated by $\sim$1--2\,pc and that the star formation is coeval within $\sim$0.1\,Myr.
\item The filamentary distributions are seen as $``$pillars$"$ similar to those in M16, extending into the Papillon Nebula. Although part of the filaments is already ionized in the H$\;${\sc ii} region, the column density of the pillars is an order of magnitude higher than that of the pillars in M16.
We suggest that these pillars were formed prior to the O-star formation, and have been continuously ionized up to the present day.
\item A recent H$\;${\sc i} study by \cite{Fukui17} suggested that kiloparsec-scale triggering of star formation occurred about 1\,Myr ago through the compression in the converging H$\;${\sc i} flows in the R136-N159 region, which was driven by the tidal interaction between the LMC and SMC. We suggest that the H$\;${\sc i} converging flows possibly triggered the coeval formation of two young star systems in the N159 region over a length scale of 50\,pc encompassing both N159E and N159W. Formation of the filaments in a cloud--cloud collision is consistent with the MHD simulations of \cite{Inoue18} which show a hub-filamentary distribution with a pivot at the O-star formation site. The present picture well explains the similar filamentary morphology of N159E and N159W, both of which show unevenly extended filaments to the north despite the large separation.  The separation does not allow communication between the two nearly coeval star-forming spots within the O-star formation timescale of $<$1\,Myr.
\end{enumerate}

\acknowledgments
We thank Doris Arzoumanian and Shu-ichiro Inutsuka for discussions about the filamentary molecular clouds.
This paper makes use of the following ALMA data: ADS/JAO.ALMA\#2012.1.00554.S, and \#2016.1.01173.S. ALMA is a partnership of the ESO, NSF, NINS, NRC, NSC, and ASIAA. The Joint ALMA Observatory is operated by the ESO, AUI/NRAO, and NAOJ. This work was supported by NAOJ ALMA Scientific Research grant No. 2016-03B and JSPS KAKENHI (grants No. 22244014, 23403001, 26247026, 18K13582, 18K13580, and 18H05440). The work of M.S. was supported by NASA under award number 80GSFC17M0002.

\appendix
\section{Appendix}
\subsection{Missing flux {\label{A:MissingF}}}
\begin{figure}[htbp]
\includegraphics[width=180mm]{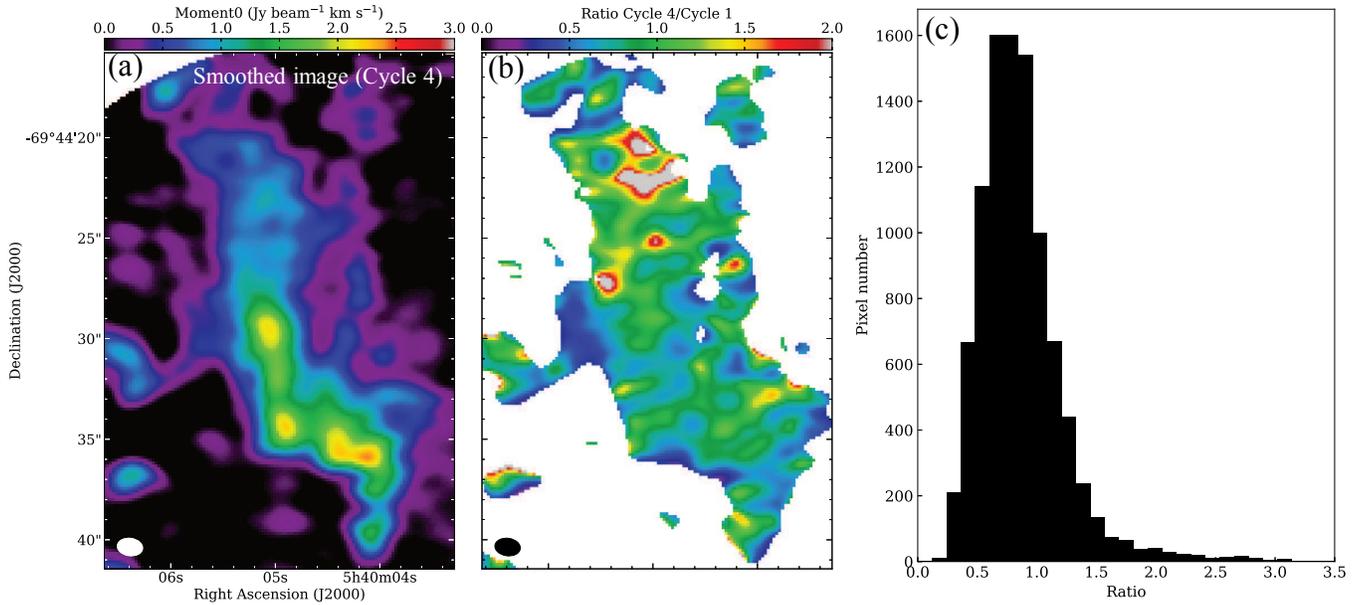}
\caption{
{Flux comparison between the Cycle 4 and Cycle 1 data in the N159E-Papillon region. (a) The smoothed velocity-integrated intensity map of the Cycle 4 $^{13}$CO\,($J$ = 2--1) data with an angular resolution of 1\farcs3\,$\times$\,0\farcs87 (the Cycle 1 data). The angular resolution is shown by ellipses in the lower left corners in panels (a) and (b). (b) The distributions of the Cycle 4/Cycle 1 intensity ratio. (c) The histogram of the Cycle 4/Cycle 1 intensity ratio shown in panel (b).
}  \label{fig:hist}
}
\end{figure}

 We estimated the missing flux of our Cycle 4 image in the N159E-Papillon region using the Cycle 1 data (Paper I). We converted the Cycle 4 data shown in Figure \ref{fig:13COfilament} (a) to be the same resolution as the Cycle 1 data (Figure \ref{fig:hist} (a)), and produced the ratio map and histogram between the two data (Figures \ref{fig:hist} (b) and (c)). The distribution of the ratio map is roughly unity and the histogram has a peak around 0.8, indicating that the Cycle 4 image has no significant missing flux compared to the Cycle 1 data. 

\end{document}